\documentclass{aa}
\usepackage{psfig}
\psfull

\def\ltsima{$\; \buildrel < \over \sim \;$}
\def\lsim{\lower.5ex\hbox{\ltsima}}
\def\loe{\lower.5ex\hbox{\ltsima}}
\def\gtsima{$\; \buildrel > \over \sim \;$}
\def\gsim{\lower.5ex\hbox{\gtsima}}
\def\goe{\lower.5ex\hbox{\gtsima}}
\newcommand{\be}{\begin{equation}}
\newcommand{\en}{\end{equation}}
\def\ergs{\rm \ erg \, s^{-1}}

\def\cmdue {\rm \ cm^{-2}}

\def\newpage{\vfill\eject}

\authorrunning{Conconi \& Campana}
\titlerunning{Optimization of surveying X--ray telescopes}

\begin{document}

\title{Optimization of grazing incidence mirrors 
and its application to surveying X--ray telescopes}

\author{Paolo Conconi\inst{1} \& Sergio Campana\inst{1}}

\institute{Osservatorio Astronomico di Brera, Via Bianchi 46, I-23807
Merate (Lc), Italy}

\offprints{S. Campana}
\mail{campana@merate.mi.astro.it}

\date{Received 15 November 2000; Accepted}

\abstract{
Grazing incidence mirrors for X--ray astronomy are usually designed in the 
parabola-hyperbola (Wolter I) configuration. This design allows for optimal images 
on-axis, which however degrade rapidly with the off-axis angle. Mirror surfaces
described by polynomia (with terms higher than order two), have been put 
forward to improve the performances over the field of view.
Here we present a refined procedure aimed at optimizing wide-field 
grazing incidence telescopes for X--ray astronomy. We improve the angular
resolution over existing (wide-field) designs by $\sim 20\%$.
We further consider the corrections for the different plate scale and focal 
plane curvature of the mirror shells, which sharpen by another $\sim 20\%$ 
the image quality. This results in a factor of $\sim 2$ reduction in the 
observing time needed to achieve the same sensitivity over existing wide-field 
designs and of $\sim 5$ over Wolter I telescopes. We demonstrate that such 
wide-field X--ray telescopes are highly advantageous for deep surveys 
of the X--ray sky.
}

\maketitle


\section{Introduction}

Imaging the X--ray sky has become possible thanks to grazing
incidence telescopes. These telescopes (the first flew on the {\it 
Einstein} observatory; VanSpeybroeck 1979; Giacconi et al. 1979) made 
possible a great improvement in sensitivity and angular resolution, 
allowing for the study of virtually all classes of X--ray sources. 
Telescopes on board {\it Einstein}, EXOSAT and ROSAT  
mainly covered the low energy band ($\lsim 4$ keV), whereas 
the production of very smooth gold surfaces made possible
the imaging up to $\sim 10$ keV with BBXRT (Serlemitsos et al. 1992), ASCA 
(Serlemitsos et al. 1995) and BeppoSAX (Citterio et al. 1991).

Two main strategies have been adopted to further develop X--ray
imaging: $i)$ increasing the angular resolution, reaching an on-axis values 
rivaling optical images ($\sim 0.5''$), as in the case of Chandra 
(Weisskopf et al. 2000); $ii)$ increasing the effective area in order to 
collect a larger number of photons (e.g. for spectral purposes) but with 
a poorer angular resolution, as in the case of Newton-XMM (Jansen et al. 2000). 
Proposed X--ray missions move along these lines: MAXIM is expected to reach 100 
microarcsec (at least; Cash et al. 2000) whereas on the other side 
{\it Constellation-X} is expected to increase the effective area ($14,500\cmdue$ 
at 1 keV) with an angular resolution of $\sim 20$ arcsec (Tananbaum et al. 1999). 
Finally, XEUS will increase dramatically the effective area (up to 
$300,000\cmdue$) with a good angular resolution of $\lsim 5''$ (Bavdaz et 
al. 1999).

X--ray astronomy benefited from serendipitous surveys (e.g. {\it Einstein} 
EMSS, Gioia et al. 1990; EXOSAT HGSC, Giommi et al. 1991; ROSAT WGA, White, Giommi 
\& Angelini 1994; ROSAT SRC, Zimmermann 1994; BMW ROSHRI, Campana et al. 1998; 
ROSHRICAT, Voges et al. 1999a; ASCA SIS, Gotthelf \& White 1997).
The first all-sky survey with an imaging X--ray telescope was carried 
out by ROSAT, spending its first 6 months surveying the entire sky. 
The ROSAT All-sky Survey (RASS, 0.1--2.4 keV; Voges et al. 1999b, 2000) is now an 
extremely useful tool for all kinds of astronomers.

As a pathfinder for new missions and to further improve the statistical studies of 
fainter/harder sources, new all-sky surveying missions are under study, such as 
ABRI\-XAS (Tr\"umper, Hasinger \& Staubert 1998) WAXS-WFXT/{\it Panoram-X} 
(Chincarini et al. 1999, 2000). 

Mirrors are usually built in the Wolter I (paraboloid-hyperboloid) configuration 
(Wolter 1952a, 1952b) which provides perfect images on-axis in principle.
This design exhibits no spherical aberration but suffers from 
field curvature, coma and astigmatism, which make the angular resolution
degrade rapidly with increasing off-axis angles (VanSpeybroeck \& Chase 1972).
Recently, Harvey \& Thompson (1999) put forward the idea of a 
telescope made by two hyperboloid surfaces (see also Nariai 1987, 
1988) which provides good performances over a field of $\sim 20'$ and 
was adopted for the Solar X--ray Imager telescope.

\begin{table}[!htb]
\caption{Grazing incidence mirror designs.}
\begin{tabular}{ccccc}
\hline
Name        & $a_2$        & $b_2$       & $a_3$ & $b_3$ \\
\hline
Parabolic   & 0            & 0           & 0     & 0 \\
Dbl.-cone &$\tan^2\alpha$&$\tan^2\beta$& 0     & 0 \\
Wolter I    & 0            & $2\,\rho_0 \,\tan \beta 
/(z_0+\rho_0\,\cot {2\,\alpha})$ & 0 & 0 \\
\hline
\end{tabular}
\label{cara}
\end{table}

\begin{figure*}[!htb]
\centerline{\psfig{figure=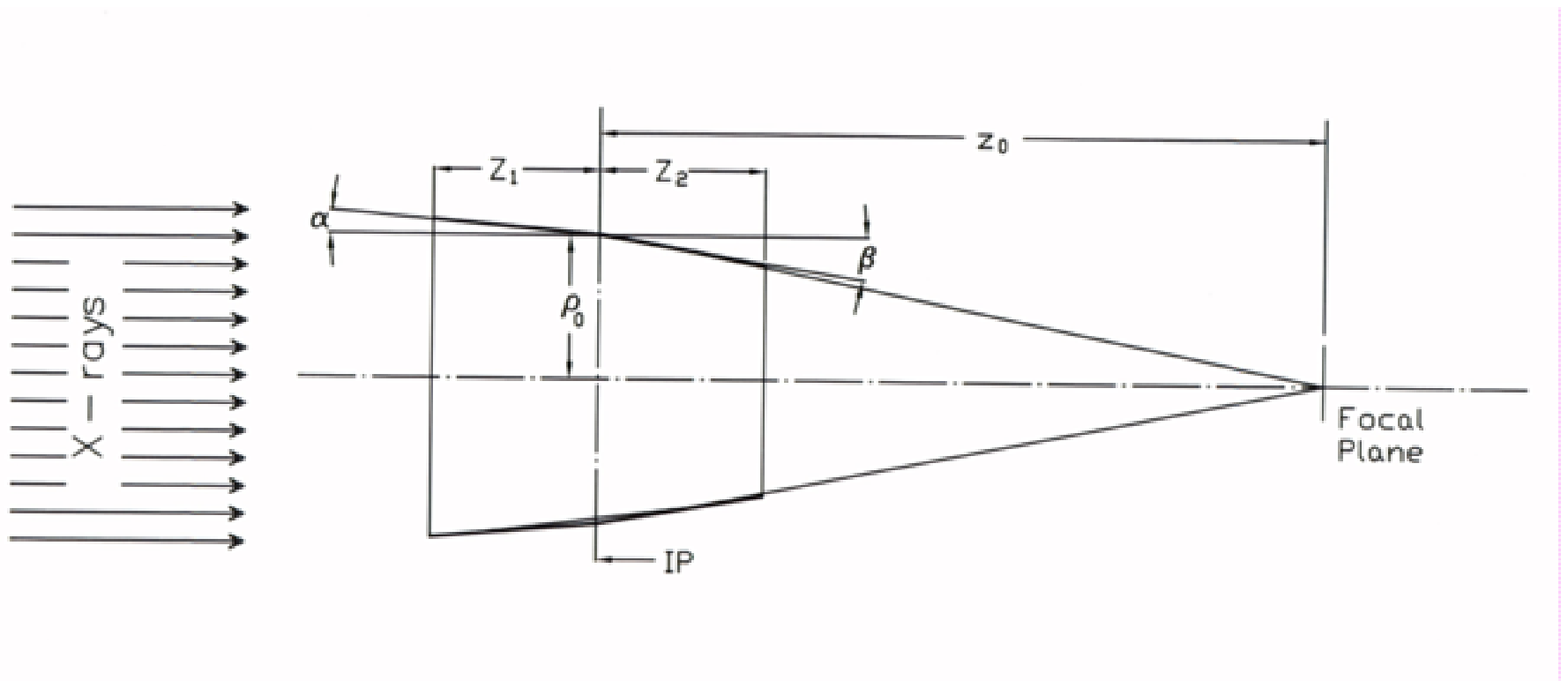,width=15cm}}
\caption{Telescope configuration and parameters.}
\label{fig_shell}
\end{figure*}

More general mirror designs than Wolter's exist in which the primary and
secondary mirror surfaces are expanded as a power series (Werner
1977; double-cone and Wolter I profiles are described by low order 
coefficients, less than two). 
These polynomial solutions are well suited for optimization
purposes, which may be used to increase the angular resolution at large
off-axis positions, degrading the on-axis performances. The idea is to transfer
the principle of the Ritchey-Chr\'etien Cassegrein telescope, widely used in 
optical astronomy, to grazing incidence optics. By deliberately compromising
the on-axis performances, one can introduce aberrations (mainly spherical) 
that tend to cancel or reduce the off-axis aberrations.

Burrows, Burg \& Giacconi (1992; BBG hereafter) discuss polynomial optics 
solutions (Werner 1977) and described in detail 
a wide-field X--ray telescope with good angular resolution up to $\sim 30'$. 
This new approach was also used to describe an optimization of the Chandra 
performances for survey work. 
Here we improve the BBG's optimization technique and sketch a procedure 
aimed at optimizing the mirror assembly. 
We further point out the need for an all-sky survey deeper and at 
energies higher than the ROSAT's. This relies on science (especially
cosmology) and on the lack of a pathfinder for the X--ray missions of the 
next generation, such as XEUS and {\it Constellation-X}.

\section{Optical Design}

The X--ray telescopes we consider here (see BBG for a more general view) 
are made by two coaxial surfaces of revolution,
which intersect at a circle, called the intersection plane (IP). A
set of parameters characterize the mirror surfaces:  the
radius of the mirror shell at the IP, $\rho_0$; the telescope focal
length, $z_0$; the angle, $\alpha$ ($\beta$), between the primary
(secondary) mirror tangent at the IP and the optical axis; the length of
the primary (secondary) mirror, $Z_1$ ($Z_2$). A shell scheme is
depicted in Fig. \ref{fig_shell}, with the origin of the Cartesian
system at the IP and the $z$ axis along the optical axis of the telescope
to the source.

In this reference system the primary and secondary mirror surfaces can be
expanded as a power series of the form 

\be
{{{\rho_1}^2}\over{{\rho_0}^2}}=\sum_{i=0}^{n_1}
a_i\,\Bigl({{z_1}\over{\rho_0}}\Bigr)^i 
{\hskip 2 truecm}
{{{\rho_2}^2}\over{{\rho_0}^2}}=\sum_{i=0}^{n_2}
b_i\,\Bigr({{z_2}\over{\rho_0}}\Bigr)^i
\label{taylor}
\en

\noindent where ($\rho_1$, $z_1$) and ($\rho_2$, $z_2$) are radial and
axial coordinates of the primary and secondary surfaces, respectively.

X--ray telescopes provide the best collecting area for a given total mirror length
(and also the best reflection efficiency at short wavelengths for a given diameter
to focal length ratio) in the case of $\xi=\alpha/(\beta-2\,\alpha)=1$ (VanSpeybroeck 
\& Chase 1972). Thus, the highest reflection efficiency 
is achieved for $\beta=3\,\alpha$ and a focal length $z_0=\rho_0/(\tan{4\,\alpha})$. 
By definition $a_0=b_0=1$, while $a_1=-2\,\tan \alpha$ and $b_1=-2\,\tan \beta$
are twice the slope of the primary and the secondary surfaces at IP. 
By selecting different values for the other coefficients one obtains different 
optical designs (see Table \ref{cara}). 

\section{Optimizing single mirror shells}

Simple recipes have been proposed and used in building X--ray telescopes to 
improve the image quality over the full field of view without affecting the 
mirror surfaces. The simplest recipe consists of slightly defocussing the 
optical system (e.g. Cash et al. 1979).
This suggestion was adopted for the flight module of the JET-X telescopes,
displacing the focal plane from the nominal value by --2.5 mm.
A different (and better) approach consists of tilting the detector so as to follow 
the (curved) focal plane. This configuration has been used for the 4 CCD detectors 
of the ACIS-I camera on board Chandra, which were assembled in an inverse shallow 
pyramid configuration. A similar approach has been adopted for the 7 CCDs of the
MOS camera: the central CCD is at the focal point on the optical axis while 
the outer six are stepped towards the mirror by 4.5 mm to
follow approximately the focal plane curvature (Turner et al. 2001).

Working directly on the mirror properties provides a more powerful tool to 
improve the image quality over the field of view.
Polynomial surfaces are particularly well suited for optimizing purposes,
since computing procedures can operate iteratively on the coefficients of 
the power series expansion. 
This can be done by defining a merit function and by finding its minimum
in the coefficient parameter space, after specifying a minimization goal.
These criteria can either provide the best image on-axis with only
a modest improvement off-axis (i.e. compromising the on-axis performances 
by a given fraction), or the flattest response over the
entire field of view (see also BBG).
In this paper we focus on the latter problem, having in mind X--ray instruments
dedicated to X--ray surveys\footnote{ROSAT mirrors have been optimized under 
a number of constraints (e.g. Wolter I geometry) to carry out an all-sky survey
(Aschenbach 1988).}. 
The key ingredients to make up an 
appropriate merit function are: $i$) an indicator of the mirror image quality at 
different off-axis positions $I(\theta)$ (e.g. spot r.m.s., Half Energy Width, 
HEW, or more generally a fixed value of the Encircled Energy Function, EEF)
and $ii$) a weighting function $W(\theta)$. If manufacturing errors and alignment 
tolerances can be summed up in quadrature to $I(\theta)$, they will not alter the 
location of the minimum.
As a general rule, the merit function can be expressed as:
\be
M={{\int_0^{\theta_0} I(\theta)\,W(\theta)\,d\theta}} \simeq
{{\sum_{i=0}^{N} I(\Theta_i)\,W(\Theta_i)}}
\en
where $N$ is a number of the fixed positions $\Theta_i$ where the image quality 
has to be evaluated. 

The minimization of the merit function is usually carried out through ray-tracing 
simulations which accounts for the mirror properties: the
image quality function $I(\theta)$ is computed for a given number of off-axis 
positions together with the weights $W(\theta)$ and the resulting merit 
function is evaluated for a specific value of the mirror parameters $\{a_j$; 
$b_j\}$. Minimization is then carried out by varying the values of the  
$\{a_j$; $b_j\}$ parameters.

As a working example, we describe the optimization of an X--ray 
telescope with a $\sim 30'$ radius field of view. 
We consider the same outer shell as in BBG (see Table \ref{mirror} for
mirror characteristics) and optimize it with the following prescriptions. 
The merit function we consider is:
\be
M=\sum_{j=0}^{n}\,\sum_{i=1}^{m}\, d_i^2(\Theta_j)\times 
[A_{\rm eff}(\Theta_j,\,E)\times W(\Theta_j)]
\en
where $n$ is the number of off-axis angles over which the merit function 
is evaluated and $m$ the number of photons used in the ray-tracing simulation 
at each position. $d_i$ is the distance between the position of the $i$-th 
photons from the center of gravity of the image at $\Theta_j$.
Unlike BBG (who adopted the spot r.m.s.), we consider a given 
fraction of the encircled energy (after positioning the spot center of gravity); 
in particular we considered the 50\% (HEW) and 80\% EEF. 
This gives a more  realistic description of the image quality in view of the 
detection of point and extended sources. 
$A_{\rm eff}$ is the effective area at a given off-axis angle (i.e. the vignetting 
function) at a given energy $E$ and allows us to optimize the mirrors where more 
photons are collected. A second weight is represented by the function $W$ defined as
\be
 W(\Theta_j)= \left\{  \begin{array} {ll} 
                         \Theta_j+\Theta_{j+1} &   0\le j < n \\
                         \Theta_n              &   j=n 
                         \end{array}
\right. 
\en
which allows us to properly weight the growing area at large off-axis angles
(in the case of equidistant sampling angles).
Finally, we consider fifth order polynomia (rather than the third order of BBG) 
and make sure that the inclusion of additional terms do not improve the 
minimization. 

\begin{table}[!htb]
\caption{WFT mirror shell characteristics.}
\small{
\begin{tabular}{cccc}
\hline
Parameter               & Outer shell & Median shell & Inner Shell \\
\hline
Focal distance          & 3000.0 mm   & 3003.3 mm    & 3005.5 mm   \\
Axial shift at IP       & 0.0 mm      & 3.4 mm       & 5.7 mm      \\
Mirror radius at IP     & 300.0 mm    & 271.1 mm     & 239.2 mm    \\
Maximum radius          & 303.0 mm    & 273.5 mm     & 241.0 mm    \\
Minimum radius          & 291.0 mm    & 263.9 mm     & 233.8 mm    \\
Mirror length$^*$       & 120 mm      & 120 mm       & 120 mm      \\
Scaled length$^*$       & 120 mm      & 100 mm       & 85.3 mm     \\
\hline
$a_2\times 10^{-3}$     & 0.87        &\ 0.69         & 0.42        \\ 
$b_2\times 10^{-3}$     & 3.99        &\ 3.31         & 2.78        \\
$a_3\times 10^{-3}$     & 2.34        &\ 2.01         & 1.39        \\
$b_3\times 10^{-3}$     & 3.40        &\ 2.53         & 0.91        \\
$a_4\times 10^{-3}$     & 3.35        &\ 3.33         & 3.11        \\
$b_4\times 10^{-3}$     &--6.45       &--4.57        &--0.81        \\
$a_5\times 10^{-3}$     & 1.47        &\ 1.68         & 2.59        \\
$b_5\times 10^{-3}$     & 5.15        &\ 3.48         & 0.04        \\
\hline
Spot r.m.s. (BBG)$^{\dag}$& 3.03      &\ 2.92         & 3.92        \\
Spot r.m.s.$^{\dag}$    & 2.76        &\ 2.59         & 2.77        \\   
\hline
\end{tabular}
}
\label{mirror}

\noindent $^*$ Lenght of the primary (or secondary) mirrors.

\noindent $^{\dag}$ Evaluated on the best tilted focal plane.
\end{table}

\begin{table}[!htb]
\caption{Mean HEW and 80\% EEF over the field of view  
for the largest mirror shell of the WFT.}
\begin{tabular}{c|cc|cc}
\hline
Mirror     &Mean     &Mean          &Mean     &Mean          \\
type       &HEW      &80\% EEF      &HEW      &80\% EEF      \\
           &(arcsec) & (arcsec)     & (arcsec)& (arcsec)     \\
\hline
Wolter I   & 7.0     & 11.4         & 7.1     & 11.6         \\
BBG        & 3.2     &\  5.2        & 4.0     &\  6.3        \\
CC$^{\dag}$& 2.7     &\  4.4        & 3.1     &\  5.4        \\
\hline
\end{tabular}
\label{sshel}

\noindent The mean has been obtained by weighting the relevant angular
response with the corresponding area of the field of view.

\noindent The first two columns on the HEW refer to the best focal position at each
off-axis angle (i.e. curved focal plane), whereas the latest two refers to the best
tilted plane.

\noindent $^{\dag}$ this work.
\end{table}

\begin{figure*}[!ht]
\centerline{\psfig{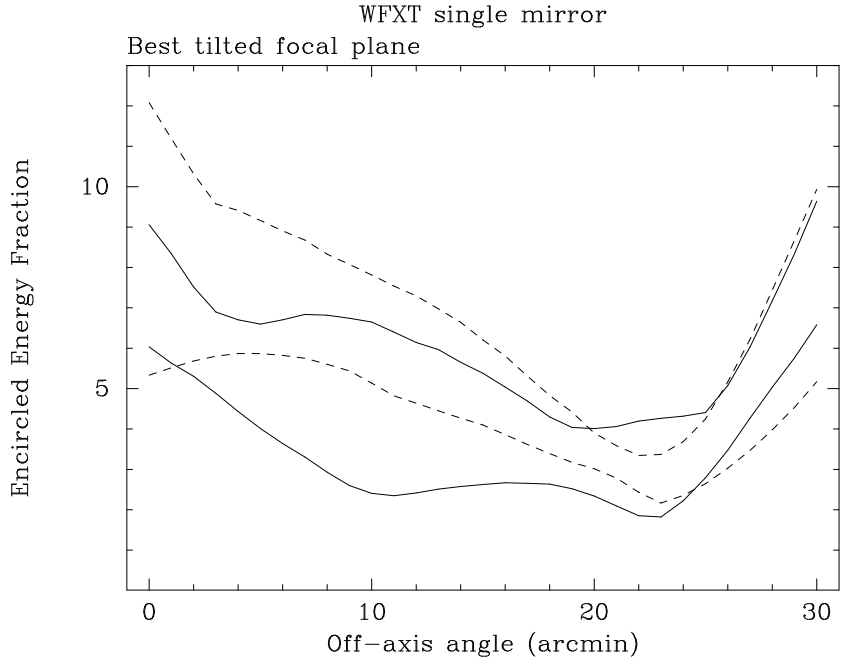}}
\caption{HEW and 80\% EEF for different mirror designs as a function of
the off-axis angle for the outermost shell of Table 2. 
The dashed lines correspond to the BBG design and the 
continuous lines to our best profile.
Of the two lines for each mirror design, the upper one refers to the 80\%
EEF and the lower to the HEW. These values are estimated at the best
focal position for each off-axis angle.}
\label{cur}
\end{figure*}

For each off-axis angle we evaluated the image characteristics at the best 
focal position, i.e. following a curved focal plane. 
In particular, we chose the curved focal plane which maximizes the 80\% 
EEF (see Fig. \ref{cur}), which is slightly different from the one individuated 
by the HEW. The profile we obtained gives better results than 
BBG's especially at small off-axis angles where the mirror effective area is 
larger and the mean HEW and 80\% EEF over the whole field of view are about 20\%
better than in the BBG design (see Table \ref{sshel}).  Because curved
CCD X--ray detectors do not exist, more realistic results can be obtained by
considering an approximation of the curved focal
plane obtained with a tilted configuration (as described for the Chandra 
satellite, see above). We evaluated the mirror
performances on the best tilted focal plane (tilt of about 3 degrees).
Also in this configuration, we obtained a mean HEW and 80\% EEF improving by a
factor of $\sim 20\%$ BBG's design (see Table \ref{sshel}).

\begin{figure*}[!htb]
\centerline{\psfig{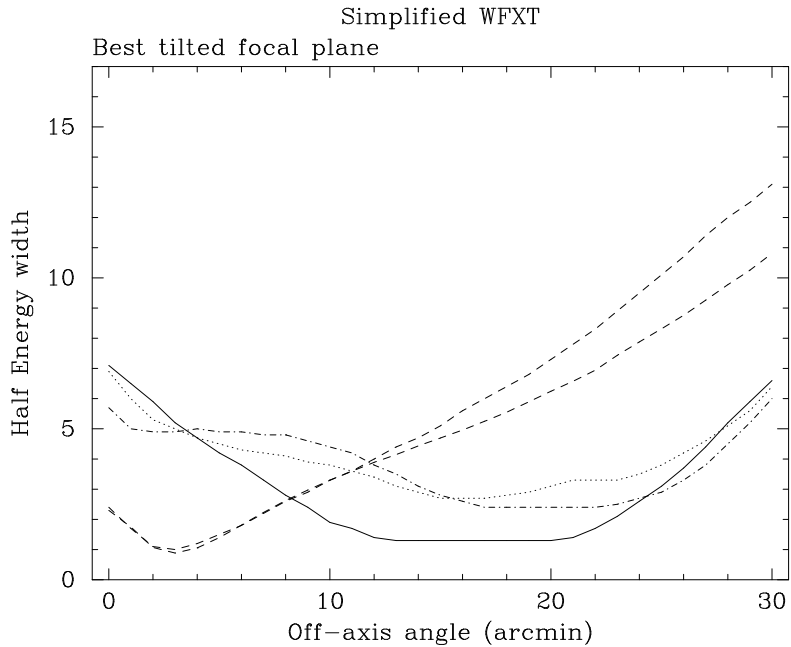}}
\caption{HEW (in arcsec) for different mirror designs as a function of the off-axis
angle, evaluated on the best tilted focal plane. The dotted line marks 
mirrors with a polynomial profile; the dot-dashed line mirrors optimized also for 
the plate scale and continuous line mirrors optimized also to have the 
same curvature of the focal plane. The dashed line represents the best 
HEW for a Wolter I telescope evaluated on the best tilted focal plane to maximize
the response over the entire field of view. The upper dashed line is without
corrections for plate scale and curvature, whereas the lower dashed line has 
these corrections included.}
\label{hew}
\end{figure*}

\begin{table}[!htb]
\caption{Image characteristics for different mirror optimizations.}
\begin{tabular}{c|ccc|cc}
\hline
Mirror    &Polyn.&Plate&Curvat.&Mean    &Mean    \\  
  type    &opt.  &opt. &opt.   & HEW    &80\% EEF\\
          &      &     &       &(arcsec)&(arcsec)\\
\hline
Wolter I  & N    &  N  &  N    &7.5     &12.3    \\
Wolter--PC& N    &  Y  &  Y    &7.0     &11.9    \\
CC--P     & Y    &  N  &  N    &3.9     &\ 7.9   \\
CC--PP    & Y    &  Y  &  N    &3.6     &\ 6.8   \\
CC--PPC   & Y    &  Y  &  Y    &2.8     &\ 5.3   \\
\hline
\end{tabular}

\noindent  The mean has been obtained by weighting the relevant angular
response with the corresponding area of the field of view.
The HEWs and 80\% EEFs have been evaluated on the best tilted 
focal plane.
\label{better}
\end{table}

\section{Optimal design for wide-field X--ray telescopes}

We consider here a simple version of a Surveying X--ray Telescope 
(SXT) consisting of three mirror shells, scaled from the largest one
described in BBG, in order to deal with the mirror assembly.
Each of these shells has been optimized separately, as described in the 
previous section.
Mirror characteristics are reported in Table \ref{mirror}.
The mean values for a confocal nesting are reported in Table \ref{better}
as CC-P. The HEW distribution as a function of the off-axis angle is plotted 
in Fig. \ref{hew} as the dotted line.

Images produced by the different mirror shells do not superpose exactly, 
having different plate scales. In particular, different shells focalize the 
relative image spots with an offset relative to one another that increases with 
the off-axis angle. 
This is a common problem of X--ray telescopes and it results in an 
increase of the image blur at large off-axis angles. This problem can be 
overcome by making the mirror shells have different intersection planes, 
i.e. shells must be moved relative to one another (see also BBG). The 
resulting HEW distribution is plotted in Fig. \ref{hew} with a dashed line 
(see CC--PP in Table \ref{better}). The relative axial shifts at the IP are 
reported in Table \ref{mirror}.

Mirrors shells of the same focal length have focal planes with different 
curvatures. This implies that for any off-axis angle the best focal plane is
calculated by averaging over the shell best focal planes. On a tilted detector  
the problem is more severe and it introduces an image blur over the entire
field of view. 
To overcome the problem one must build mirror shells which have 
the same best focal plane (i.e the same curvature). 
This can be achieved by constructing scaled versions of the outermost shell, 
implying that the innermost shells are shorter. 
The reduction of the inner shells results in a smaller effective area. 
This can be compensated by making the starting outermost shell longer.

\begin{table*}[!htb]
\caption{All-sky surveys with existing and planned X--ray telescopes.}
\begin{tabular}{ccc|c}
\hline
                                  & ROSAT    & ABRIXAS  &  SXT     \\
\hline
Energy range (keV)                &0.1--2.4  & 0.3--10  & 0.3--10  \\
Area @ 1.5 keV (cm$^2$)           & 200      &  80      &  600     \\
HEW on-axis (arcsec)              &  20      &  20      &   10     \\ 
Average HEW over field of view    &  60      &  40      &   10     \\
Solid angle (deg$^2$)             &  3       &  2.4     &  1.4     \\
Average exposure time (s)         &  500     & 4,000    & 2,000    \\
Time for all-sky (yr)             & 0.5      &   3      &   3.0    \\ 
Indicative positional accuracy ($1\,\sigma$)$^\dag$ &  $25''$  &  $12''$ &    $3''$   \\  
\hline
Limiting flux (cgs) (0.5--2 keV)  &$5\times10^{-13}$&$1\times10^{-13}$&$1\times10^{-14}$\\
Limiting flux (cgs) (2--10 keV)   & --              &$4\times10^{-13}$&$1\times10^{-13}$\\
\hline
\end{tabular}
\label{year}

\noindent 
For SXT we consider 50 mirror shells reaching an effective area of 600 cm$^2$ 
at 1.5 keV (after convolution with the CCD and filter responses).
The sensitivity has been computed requiring that at least 5 photons are 
collected within the detection cell and adopting the Chandra background 
of 0.3 ct s$^{-1}$ chip$^{-1} = 3\times 10^{-7}$ ct s$^{-1}$ pixel$^{-1}$. 

\noindent 
$^\dag$ Accuracy of the Point Speread Function centroid position at
the limiting flux level.

\end{table*}

Table \ref{better} and Fig. \ref{hew} show the improved performance
of our simple SXT. The polynomial design improves the mean HEW over the 
$30'$ field of view by about 100\% over the Wolter I design. 
The plate correction results in a small improvement ($\sim 10\%$), and the 
plate and curvature optimization (CC--PPC in Table \ref{better})
provide a gain of $\sim 40\%$ over the 
polynomial design. In particular, the curvature optimization is more important 
and mirror performances are improved especially in the $10'-20'$ region 
(see Fig. \ref{hew}, continuous line). 

\begin{figure*}[!htb]
\centerline{\psfig{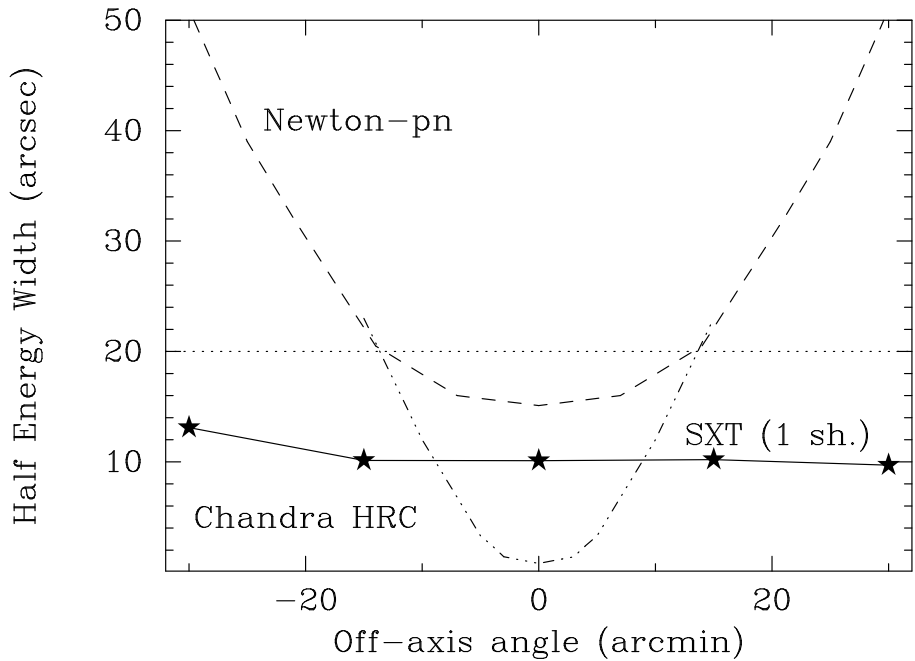}}
\caption{Comparison of the HEW of the SXT mirror shell realized by Citterio et al.
(1998; stars and continuous line) with current X--ray telescopes vs. the off-axis 
angle. A dot-dashed line marks Chandra-HRC and a dashed line Newton-pn instruments.
The SXT HEW over the off-axis angle does not appear symmetric over the field 
due to a $-5'$ mis-pointing in the calibrations (i.e. an off-axis angle of $-30'$ 
corresponds to $-35'$). A horizontal dotted line represents the
$20''$ limit adopted in the text. Adapted from Citterio et al. (1998).}
\label{sate}
\end{figure*}

\section{Need for a wide-field telescope for all sky surveys}

In the infancy of X--ray astronomy several collimated experiments 
surveyed the X--ray sky (e.g. {\it Uhuru}, OSO-7 and HEAO-1).
The first imaging all-sky survey was conducted by ROSAT.
The RASS has been extremely successful and now represents 
a powerful tool for X--ray astronomy (Voges et al. 1999b, 2000).
The RASS revealed 18,881 sources with a count rate higher than 0.05 c s$^{-1}$ 
(comprising the Bright Source Catalog, BSC; Voges et al. 1999b) and 105,924 
sources with a likelihood ratio higher than 7 (Voges et al. 2000). Due to the 
scanning of great circles, the exposure times are highly variable from the 
equator to the ecliptic poles. A mean exposure time of 500 s can be considered 
(Voges et al. 1999b). Assuming a limit of 0.05 c s$^{-1}$ for the survey 
(i.e. the same of the BSC) a rough flux limit of $5\times 10^{-13}\ergs\cmdue$ 
(0.1--2.4 keV) can be derived.
The Position Sensitive Proportional Counter (PSPC) detector with which 
the RASS has been carried out allowed for a HEW of $\sim 20''$ on-axis 
and of $160''$ at 50 arcmin off-axis. A mean HEW of about $60''$
is expected for a scanining trace passing from the center (and worse for the 
others). The resulting positional accuracy is for sources in the BSC 
at a level of $25''$ (90\% confidence level; Voges et al. 1999b).

The need for an all-sky survey at high energies (0.5--10 keV) led to the 
development of the ABRIXAS mission which unfortunately failed. The optical 
system is a bundle of seven Wolter I telescopes (each consisting of 27 nested 
gold-coated mirror shells) titled with respect to each other, focussing images on 
a single pn CCD (Tr\"umper, Hasinger \& Staubert 1998; Predehl 1999). 
The on-axis HEW is about $20''$ (Predehl 1999) degrading to $\sim 80''$
at $20'$. The resulting mean resolution is about $40''$. A rough positional 
accuracy estimate can be derived by the ratio ${\rm HEW}/(2\,\sqrt{N-1})$ (e.g. 
Lazzati et al. 1998, since for a 2-dimensional Gaussian the HEW coincides 
with the $2\,\sigma$) where $N$ the number of photons. So for a 10 photons
detection limit a positional accuracy of $\sim 12''$ is achieved (90\% confidence
level).

The main driver of the ABRIXAS mission and in general of sky surveys at 
energies higher than 2 keV was the study of the extragalactic background and 
of absorbed Active Galactic Nuclei (AGN). Moreover, clusters of galaxies are 
among the best tracers of the large-scale structure and their evolution 
is a powerful diagnostic for the geometry of the universe (e.g. Borgani \& Guzzo 
2001). Clusters at any redshift and cosmology can be detected as extended 
sources if the HEW is better than $\sim 15''$. 
In a flux-limited survey down 
to $\sim 10^{-14}$ erg s$^{-1}$ cm$^{-2}$, clusters of galaxies will be detected 
in large numbers up to $z\sim 1$ and AGN up to $z\sim 4$. 
With that good angular resolution, clusters of galaxies can be detected 
and picked up {\it directly} in the X--ray images as extended sources.

A more efficient way to carry out the mapping of the X--ray sky is to use an 
optimized mirror system with a large corrected field of view.
We consider here a Surveying X--ray Telescope (SXT) made by optimized mirrors
with an effective area comparable to that of the Chandra instrument, 
i.e. $A_{\rm eff} \sim 600$ cm$^2$ at 1.5 keV. 
This number includes the mirror effective area, the CCD quantum efficiency and 
filter transmission. We consider here CCDs like the EPIC/MOS 
on board XMM-Newton or the ACIS-I on board Chandra. These are characterized by a 
small pixel size but by a reduced response at high ($>4$ keV) energies, at variance 
with the EPIC-pn CCD adopted by ABRIXAS.
Iridium coating of the mirrors will improve the response
of the SXT at high energies providing a total 
effective area at 6 keV of $\sim 130$ cm$^2$.
For an outermost shell diameter of 70 cm diameter, we would need 50 mirror shells
to reach the selected area.
The corrected field with an HEW $<20''$ extends up to $\sim 40'$.
In order to cover the focal plane, 9 CCDs are needed.
These should be displaced in a way similar to the EPIC-MOS CCDs (see above)
in order to follow the curved focal plane.

As a second parameter we set the time to carry out an all-sky 
survey as 3 years. Assuming a field of view of $40'$ radius and allowing 
for a 20\% superposition of different scans we end up with a 2,000 s 
mean observing time per field.
For the computation of the survey sensitivity we assume the background
of the Chandra ACIS-I ($3\times 10^{-7}$ ct s$^{-1}$ pixel$^{-1}$),
an absorbed power law (photon index 2 and column density of $3\times 
10^{20}\cmdue$) for the source spectrum and a source detection with 5 photons 
in the detection cell. 
This last condition is motivated by the very low background in the small 
detection cell ($\lsim 0.09$ counts in 2,000 s per detection cell), resulting 
in a $\sim 5.3\,\sigma$ detection. With these constraints we obtain a limiting 
0.3--10 keV flux of $2\times 10^{-14}\ergs\cmdue$. The flux limit reduces to 
$1\times 10^{-14}$ and $1\times 10^{-13}\ergs\cmdue$ in the 0.5--2 and 
2--10 keV energy bands, respectively (see Table \ref{year}). An indicative 
positional accuracy at the detection limit of $\lsim 5''$ (90\% c.l.) 
can be reached.

Such a survey would be the analogue of the Palomar survey in the X--ray and 
would be useful for the next generation of X--ray satellites, XEUS and 
{\it Constellation-X}, as well as to pinpoint places where NGST could 
image the earliest phases of the forming structure in the Universe.

\section{Conclusions}

X--ray surveys can be more efficiently carried out with mirrors
optimized over a large field of view. The polynomial design is superior with 
respect to the Wolter I design.
We show here that our design provides an improvement in the mean HEW
over the entire field of view of $\gsim 150\%$ over the Wolter I design. This
makes it possible to carry out sky surveys in a shorter time and/or at deeper levels.

In this paper we describe a refinement of the optimization technique 
put forward by BBG and Werner (1977). We are able to improve the 
design of single mirror shells for wide-field imaging by about 40\%
in terms of mean angular resolution (HEW) with respect to the original 
BBG's design, i.e. source detections at the same significance level can be achieved 
in about a factor of 2 shorter exposure time. 

We compare the performances of a wide-field optimized telescope (SXT)
sized to an effective collecting area similar to Chandra, 
with existing and planned mission surveys. The SXT gains a factor of 
$\sim 10$ in sensitivity mainly due to the improved angular resolution.
Furthermore, current and planned X--ray telescopes fail to provide a deep mapping 
of the X--ray sky during slews whereas serendipituous surveys have access to only 
a few percent of the sky. In particular, either Chandra and XMM-Newton fail 
to provide an all-sky survey at the selected limiting flux in a reasonable time 
($\gsim 20$ yr, if pointings of 2 ks are carried out).

Technology requirements are severe for the manufacturing of these optimized shells, 
even if the experience of Chandra has indicated that mirrors with tiny
mechanical tolerances can be made (paying however for a large mirror weight). 
Despite these caveats, a mirror shell following an earlier optimized 
design has already been built (Citterio et al. 1998, 1999).
These authors have demonstrated that, using an upgraded replication technique, 
it is possible to produce light-weight carriers in ceramic material (Silicon 
Carbide, SiC, or Allumina, Al$_2$O$_3$; instead of nickel carriers used for 
BeppoSAX, JET-X, Newton-XMM and Swift missions), over which the reflecting 
material (gold in this case) can be deposited. 
As a result of this process, a single mirror shell with a HEW $< 13''$ over a 
$35'$ (radius) field of view has been built and tested (see Fig. \ref{sate}; 
Citterio et al. 1999; Ghigo et al. 1999).
This demonstrates that current technologies meet the fabrication requirements 
(especially stiffness, e.g. Citterio et al. 1998) of these optimized 
mirror designs. 

Comparison with current and future X--ray missions shows that {\it only} a 
dedicated mission with an optimized mirror is able to produce in a reasonable time 
an all-sky survey down to a limiting flux of a few $10^{-14}\ergs\cmdue$ 
in the 0.5--10 keV energy band.

\begin{acknowledgements}
We acknowledge useful discussions with G. Chincarini, G. Ghisellini,
G. Pareschi \& G. Tagliaferri. 
This work was partially supported through ASI grants.
We thank the referee (B. Aschenbach) for his comments which improved and
sharpened the paper.
\end{acknowledgements}

\end{document}